\documentclass[12pt]{iopart}
\usepackage{graphicx}
\def\WIDTHA{\textwidth}  %% 1 FIGURE WIDTH    / PREPRINT
\begin{document}

\letter{What is the length of a knot in a polymer ?}

\author{B. Marcone, E. Orlandini, A. L. Stella and F. Zonta}
\address{INFM-Dipartimento di Fisica and Sezione INFN,Universit\`a di Padova,
I-35131 Padova, Italy.}
\ead{orlandini@pd.infn.it}

\begin{abstract}
We give statistical definitions of the length, $l$, of
a loose prime knot tied into a long, fluctuating ring macromolecule.
Monte Carlo results for the equilibrium, good solvent regime show that
$\langle l \rangle \sim N^t$, where $N$ is the ring length
and $t \simeq 0.75$ is independent of the
knot type. In the collapsed regime below the theta 
temperature, length determinations based on the entropic
competition of different knots within the same ring
show knot delocalization ($t \simeq 1$).

\end{abstract}
\submitto{\JPA}
\maketitle
Long polymers are remarkable systems in which topological entanglement
is almost unavoidable~\cite{SW,Muthu} and
shows consequences pertaining to many different scientific contexts.
In chemistry the synthesis of macromolecules with nontrivial
topology, like catenanes, is well established~\cite{FW}. The role of knots
and links in the biology of DNA and the existence of enzymes like
topoisomerases, which control these entanglements,
have been at the basis of an explosion of interest in topology
among biologists, chemists, mathematicians, and physicists since the
eighties~\cite{WC1,Sumners}.
Most recently, techniques for manipulating
single molecules opened new possibilities of knot 
creation and study~\cite{Arai99,Quake03}. 

In all these examples the knowledge of the degree of localization
of the knots within the chains is of primary importance. 
The action of topoisomerases~\cite{Sumners,WC1}
certainly depends on how localized the 
topological entanglement is~\cite{Yan99}. 
The folding dynamics of a knotted protein~\cite{Taylor00}
should also be strongly influenced by the tightness of the knot.
When a knot diffuses along a DNA molecule
stretched by molecular tweezers, the
length of the entangled region is also essential
since it determines the diffusion coefficient. 
If the knot is tight enough this length can be approximated by that
of the knot in its ideal form~\cite{Quake03}.
An ideal knot is realized with the shortest piece of rope
of constant diameter and is the only one for which a precise definition
of length exists~\cite{Katrich96,Cantarella,Buck}.
Of course, this model is inadequate if the molecule is not well stretched
and the knotted region fluctuates substantially. 

In spite of several indications that
prime knots of polymers in good solvent are rather well localized
~\cite{OTJW98,KOVDS00},
proofs and direct quantitative descriptions of this 
localization are still missing.
This is primarily due to difficulties
related to the notion of a knotted open string~\cite{RSWW}.
Indeed, to localize the knot requires to identify an open portion of the
ring which is ``knotted".
However, to assert that an open piece of a closed rope
``contains" a knot, is ambiguous.
Mathematically there are no knots in an open string
since continuous deformations can always
bring it into an untangled shape.
Substantial progress
in the description of localization properties of topological
entanglement has been made for flat knots~\cite{Guitter}.
These models describe three-dimensional knotted
polymers fully adsorbed on a plane with the simplifying
feature that the number of overlaps is restricted
to the minimum compatible with the topology. In this way knot length 
can be defined and studied both numerically 
and analytically~\cite{Kardar}.
It was also discovered that flat knots become delocalized
when the adsorbed polymer undergoes theta collapse~\cite{OSV03,Kardar03}. 
A demonstration of the conjectured~\cite{OSV03}
analogous delocalization phenomenon for compact knotted
polymers in three dimensions remains an open challenge~\cite{Grosberg04}.

In this Letter we propose a statistical description of the lenght, $l$, of
a knot within a circular fluctuating polymers with $N$ monomers.
We find that $\langle l\rangle\sim N^t$ with $t \simeq 0.75$
in the good solvent regime (weak localization), and $t\simeq 1$
(delocalization) in the collapsed phase~\cite{Vanderzande}.

We consider knotted self-avoiding
polygons (SAP) on cubic lattice, which
are a good model for long polymers in a
good solvent~\cite{Vanderzande}.
If necessary, a theta collapse~\cite{Vanderzande} into compact configurations
can be induced by switching on an attractive interaction
between nearest neighbor lattice sites visited not consecutively
by the SAP~\cite{OSV03} and by lowering enough the temperature $T$. 
$N$ is the number of lattice edges occupied by the SAP.
A BFACF Monte Carlo dynamics~\cite{CPS} preserving the topology 
of the SAP~\cite{JW91b} enabled  the sampling of equilibrium configurations of
chains of variable $N$.
To increase mobility a multiple Markov chains~\cite{TJOW} in the 
space of the edge fugacity  was implemented.

A first strategy of knot length determination
is as follows.
In each sampled configuration, various open portions of
the SAP are considered and for each one of them 
a closure is made by joining
its ends, A and B, with an off-lattice path (Fig.1). 
%%%%%%%% FIG  1  %%%%%%%%%%%%%%%%%%%%%%%%%%%%%%%%%%%%%%%%%%%%%%%%%
\begin{figure}[!bp]
\includegraphics[width=\WIDTHA]{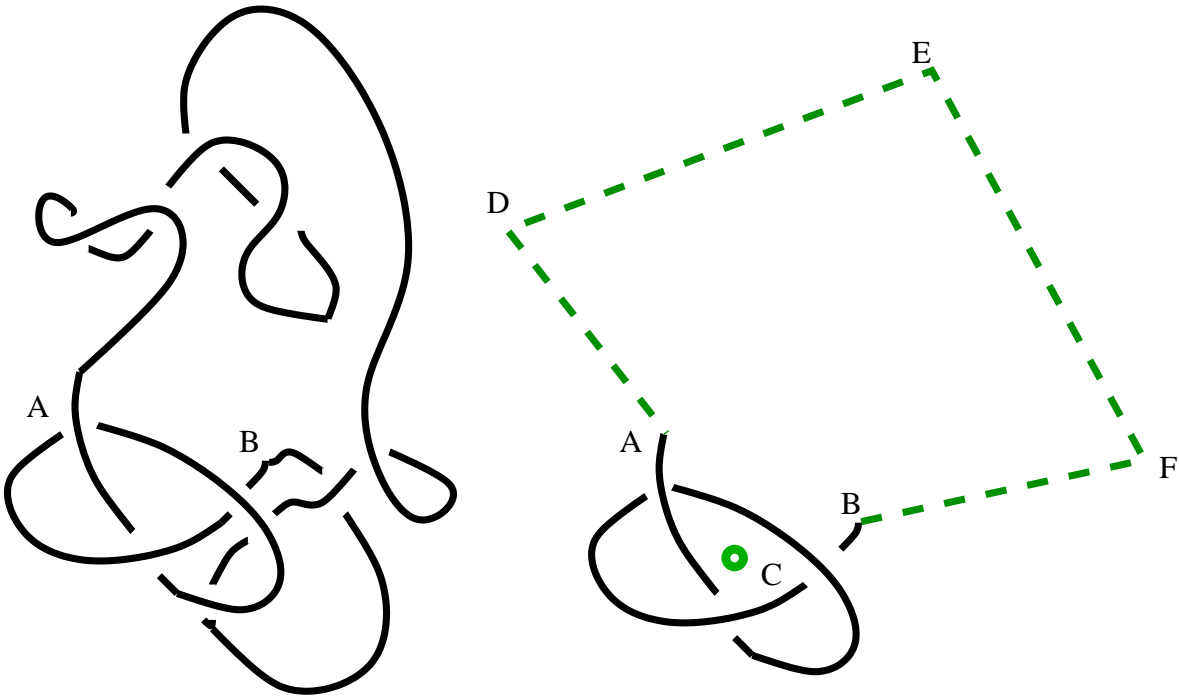}
\caption{ \label{fig1}
Once extracted the open string portion (AB) from the SAP, we  
determine its center of mass C.
An off-lattice planar 4-edges polygonal is attached to the  
extrema A and B.
The polygonal lies on the plane  CAB.
D,E and F are all chosen far
away from C in order to avoid additional entanglement with the open
string.
}
\end{figure}
%%%%%%%% FIG  1  %%%%%%%%%%%%%%%%%%%%%%%%%%%%%%%%%%%%%%%%%%%%%%%%%%%%%%%%%^M
This path is chosen in order to minimize
the risk of  knot modifications or disentanglements
in the resulting new ring. This risk can not be
fully avoided and is a possible source of  systematic errors. 
Once the new ring has been constructed, the
computation of a topological invariant, namely the Alexander 
polynomial~\cite{Rolfsen} $\Delta(z)$ at $z=-1$, allows to
verify  the presence and the type of  knot~\cite{note0}. 
The length $l$ of the knot in
a given SAP configuration can then be identified with the shortest
ring portion still displaying the original knot, among
a large set obtained by several cutting and closing operations~\cite{nota}.
The plot of the average length of a trefoil knot ($3_1$)~\cite{Rolfsen}
as a function of $N$ (Fig.2) shows full consistency with the scaling
law
\begin{equation}
\langle l \rangle \sim N^t
\end{equation}
with $t = 0.74 \pm 0.14$ which  is also robust
with respect to a change of prime knot type. For example, if we replace 
the $3_1$ by a $4_1$ or $5_1$ knot~\cite{Rolfsen}, Eq.(1) remains valid with the same $t$
within confidence limits. Unfortunately, one can  not estimate
the possible systematic errors arising from the cutting
and closing procedure used in this analysis~\cite{note1}, which 
would be a serious handicap for applications to compact 
polymers.
Thus,  one needs to validate the law in Eq.(1) by alternative, 
consistent methods of length determination.

%%%%%%%% FIG  2  %%%%%%%%%%%%%%%%%%%%%%%%%%%%%%%%%%%%%%%%%%%%%%%%%%%%%%%%%^M
\begin{figure}[!tbp]
\includegraphics[angle=0,width=\WIDTHA]{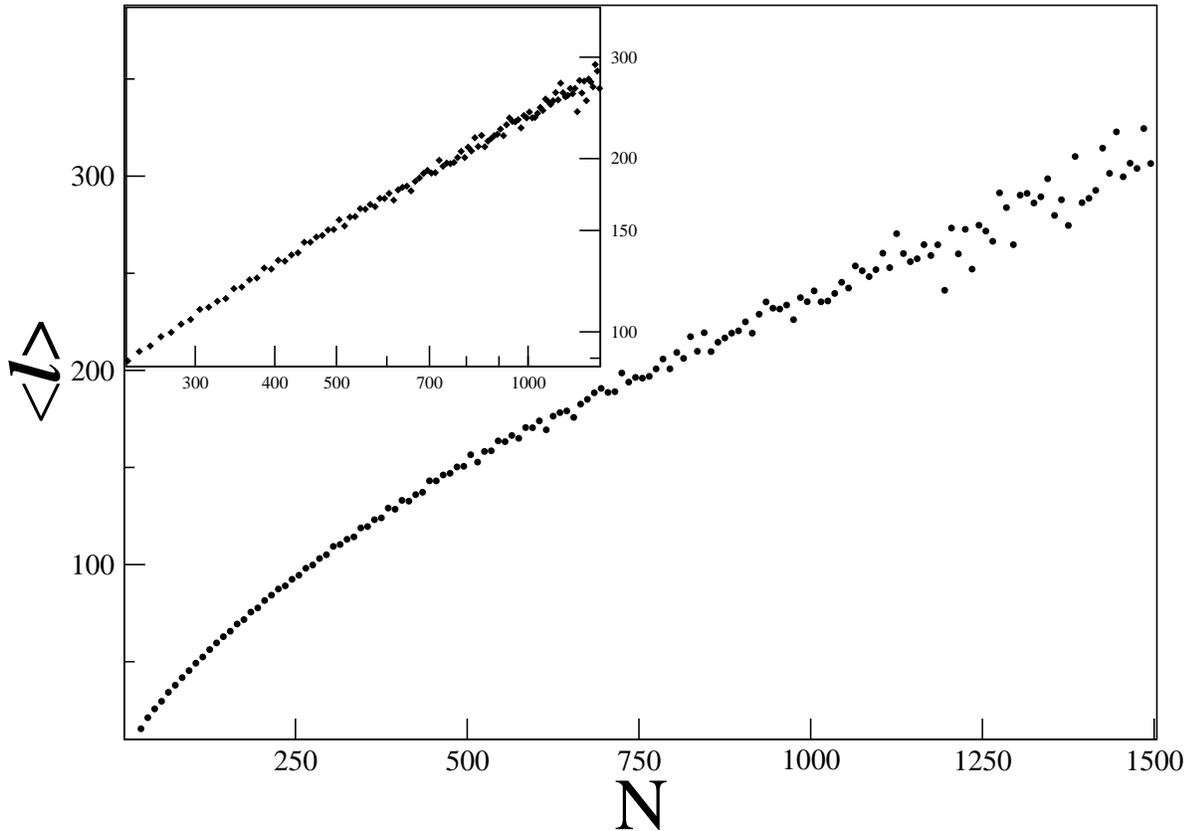}
\caption{ \label{fig2}
$\langle l \rangle$ of the trefoil ($3_1$) knot as a 
fuction of $N$. 
The configurations sampled were $\sim 10^6$ in total. 
The brackets $\big< . \big>$ denote averages taken over data binned
around fixed values of $N$.
The inset shows the log-log version of the  plot.
}
\end{figure}
%%%%%%%% FIG  2  %%%%%%%%%%%%%%%%%%%%%%%%%%%%%%%%%%%%%%%%%%%%%%%%%%%%%%%%%^M

Consider a SAP partitioned into two
loops by a narrow slip link. Each loop is tied into a $3_1$ knot and the
Monte Carlo dynamics is such that each knot cannot translocate
from its loop to the other one (Fig. 3). 
The two loops also remain unlinked.
%%%%%%% FIG  3  %%%%%%%%%%%%%%%%%%%%%%%%%%%%%%%%%%%%%%%%%%%%%%%%%%%%%%%%%
\begin{figure}[!tbp]
\includegraphics[angle=0,width=\WIDTHA]{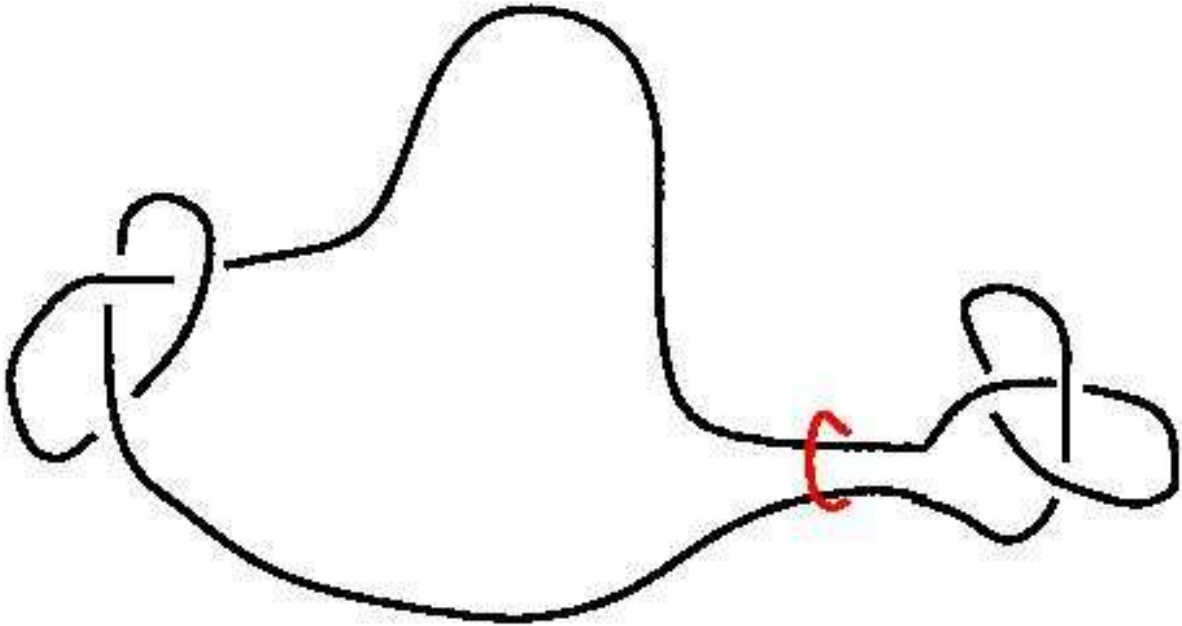}
\caption{ \label{fig3}
Sketch of a $3_1$ knot (right) forced to its typical length by the
entropic competition between knotted loops. 
To simulate the slip link we impose the constraint 
that two fixed parallel edges of the polygon have always to be kept one 
lattice distance apart~\cite{nota}. 
}
\end{figure}
%%%%%%%% FIG  3  %%%%%%%%%%%%%%%%%%%%%%%%%%%%%%%%%%%%%%%%%%%%%%%%%%%%%%%%%^M
Since the number of configurations for the 
whole ring is maximum when one of the loops is much longer than the other one,
most configurations break the symmetry between the two
loops showing a marked length unbalance.
Typically, in one of the two loops the knot has a very
large share of the whole ring at its disposal, while the other loop is just 
long enough to host its knot. Consenquently, we choose to always identify
the length of the shorter loop with the length, $l$, of a trefoil
knot inserted in a SAP.
Such $l$ is always sampled now in situations in which
the two ends of the knot come close to each other in the neighborhood of the
slip link. However, one can hope this statistic 
to be representative of more general samples if laws like
Eq.(1) are considered. 
Indeed, for the average of this $l$ the
power law behaviour reported in Eq.(1) still
holds, with $t = 0.74 \pm 0.05$, consistent
with the previous estimate (Fig. 4). 
In the case of two unknotted loops we verified that 
the average length of the shorter one does not
grow appreciably with $N$, further supporting our interpretation of $l$ as the
knot length in the $3_1$ vs $3_1$ case.

%%%%%%%% FIG  4  %%%%%%%%%%%%%%%%%%%%%%%%%%%%%%%%%%%%%%%%%%%%%%%%%%%%%%%%%
\begin{figure}[!tbp]
\includegraphics[angle=0,width=\WIDTHA]{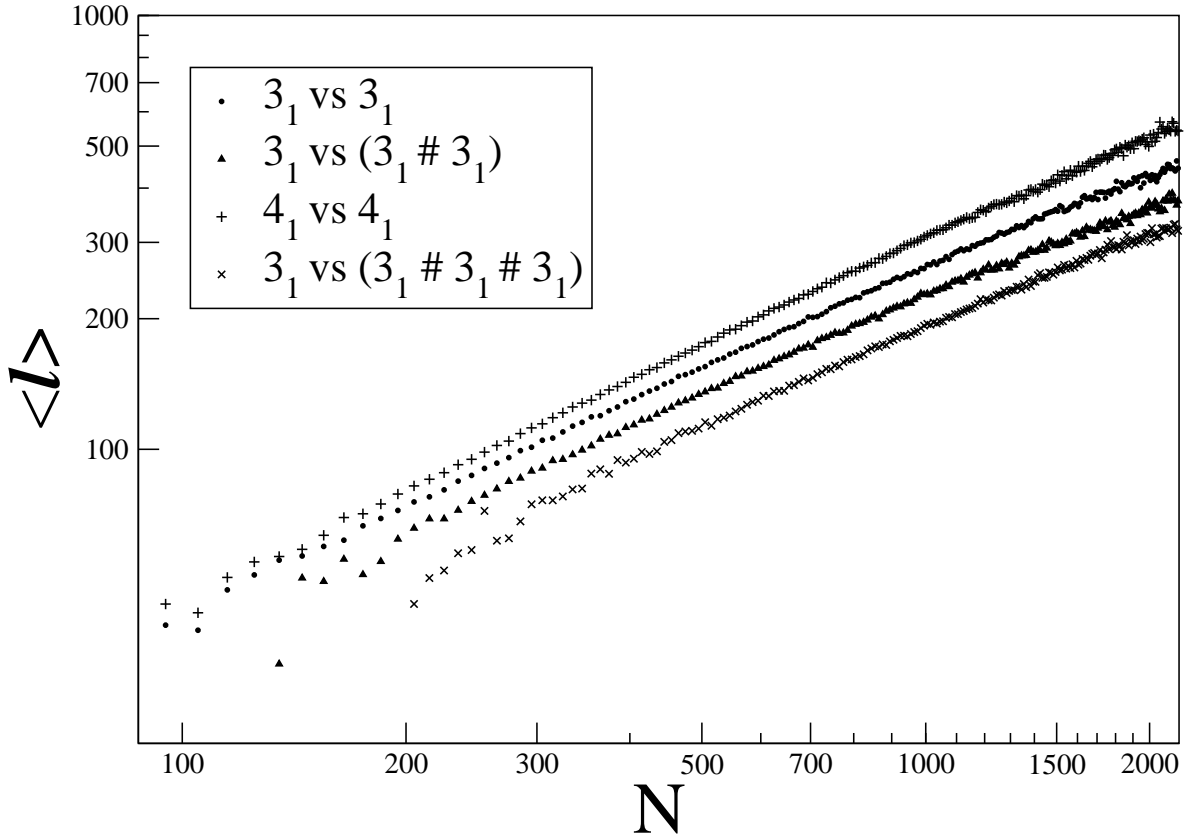}
\caption{ \label{fig4}
Log-log plot of the average value of the length $l$  of the shortest
loop as a function of $N$. 
Different curves correspond to the presence of different knots
in the two loops. 
Best fit estimates of $t$ are
$0.74 \pm 0.05$, $0.75\pm 0.03$, $0.77\pm 0.10$, $0.74\pm 0.05$ 
for $3_1 \hbox{vs} 3_1$,  
$3_1 \hbox{vs} 3_1 \# 3_1$, $4_1 \hbox{vs} 4_1$ 
and $3_1 \hbox{vs} 3_1 \# 3_1 \# 3_1$, respectively.
}
\end{figure}
%%%%%%%% FIG  4  %%%%%%%%%%%%%%%%%%%%%%%%%%%%%%%%%%%%%%%%%%%%%%%%%%%%%%%%%^M

This last method of length determination needs 
a suitable competition between the two loops~\cite{Metzler03}.
If only one of them were left unknotted,
a too strong  dominance of the knotted loop would result,
making it almost impossible to sample configurations
in which the longest loop is the unknotted one. 
Indeed, since the knot is rather localized,
its freedom to place itself along the loop multiplies
by a factor roughly proportional
to the length of the loop itself the number of configurations
accessible in its absence. 
By putting  a $4_1$ knot against another $4_1$ knot, within error bars,
we found the same value of $t$ (Fig. 4).
Thus, like in the previous strategy, $t$ appears universal
with respect to the prime knot type.
We also carried out  simulations in which one loop contained a 
trefoil, while the other
one was tied into a composite knot made, e.g., by the product
$3_1 \# 3_1$ of two trefoils~\cite{Rolfsen}. In this case the loop hosting
the composite knot almost constantly dominates in length because both
components are rather localized and free to move independently along it.
In spite of the different conditions of entropic balance, 
we verified that also in this case the average length of the 
(shorter) loop with a single knot satisfies Eq.(1) with the same 
$t\sim 0.75$ (Fig. 4). 

The above results have implications for
the leading scaling correction to the
singular behavior of quantities like the average radius of gyration of
the SAP, $\langle R_g^2 \rangle^{1/2}
\sim A N^{\nu} (1 + B N^{-\Delta})$. Field theoretical renormalization group
(RG) methods~\cite{Zinn} and
most Monte Carlo techniques~\cite{Li} allow to determine exponents like
$\nu\simeq 0.588$ and $\Delta \sim 0.50$ in ensembles in which 
configurations with all possible ring topologies are sampled.
$\nu$ is believed to remain the same in ensembles like those
considered here, with fixed knot topology~\cite{OTJW98}.
With our first cutting and closing method, 
we could check that the ring portion occupied by the knot
has an average radius of gyration $\sim l^{\nu_0}$, with 
$\nu_0 \simeq 0.60 \simeq \nu$.
This means that
a length diverging as $\langle R_0 \rangle \sim N^{\nu t}$
exists in addition to the leading one $\langle R_g^2 \rangle^{1/2}
\sim N^{\nu}$. Standard  scaling arguments then lead to
expect a possible strong correction exponent $\Delta =1-t \sim 0.25$ for
the ensemble with fixed knot topology. 
Such stronger correction could be quite difficult to identify 
numerically at relatively short $N$~\cite{DDMSS03}. 

Our strategy based on entropic competition has the advantage of being
applicable also to situations in which the polymer
is compact. Indeed, since this method
does not involve cutting and closing procedures,
there is no risk that the knot topology gets altered,
a highly probable event for compact configurations.
This enabled us to address another key open issue
in the field, namely the possible occurrence
of knot delocalization in the compact regime~\cite{OSV03}.
We included nearest neighbor attractive interactions
for the two loop model and simulated it extensively
at $T\sim 0.53 T_{\theta}$, where $T_{\theta}$
is the theta temperature~\cite{Vanderzande,TJOW}.
Below the theta point it becomes more difficult
to sample long SAP configurations. Indeed, at $T<T_{\theta}$,
as the critical edge fugacity is approached
from below, the grand-canonical average number of
SAP edges, $\langle N \rangle_{g.c.}$, undergoes a
first order infinite jump from a finite value,
rather than growing continuously to infinity as
in the $T>T_{\Theta}$ case.
In spite of this difficulty, we could get rather clear
evidence that the canonical average $\langle l \rangle$ 
grows linearly with $N$ up to $N\sim 700$, implying
$t\simeq 1$, for
the case of two competing $3_1$ knots (Fig. 5). 
Like when dealing with the good solvent regime,
in order to interpret the length of the shorter loop 
as the knot length, it is important to check what
happens to the length of a loop without knot.
We got evidence that in the compact regime an unknotted loop
is also delocalized, i.e. shows an average length
proportional to $N$. However, the proportionality
constant is about one order of magnitude smaller
than that observed for the knotted loop
in Fig. 5. 
Thus, knot delocalization seems to superimpose itself
to a phenomenon of unknotted loop delocalization
which is qualitatively similar, although quantitatively 
much less pronounced. The entropic competition
between loops in the compact phase does not lead to
a strong prominence of one of them on the other, like
at high $T$. To the contrary, all loop length
ratios are realized with comparable probability.
In this new scenario also the entropic mechanisms
determining knot localization in the good solvent case
cease to be active. 

The delocalization of prime knots in 
compact polymer rings opens intriguing 
perspectives on how topological
entanglement combines with more geometrical measures
of complexity in polymers. One such measure for
a fluctuating SAP is the mean number
of crossings, $N_c$, one obtains by projecting
on all possible planes a given configuration~\cite{RSWW}.
This quantity can be studied also for
ideal knots~\cite{Katrich96}, and has been shown to be correlated with the
mobility under electrophoresis of non-ideal knotted
polymer rings with the same topology~\cite{Stasiak96}.
For ideal knots, $N_c$ is expected to grow as
$l_{id}^{4/3}$, where $l_{id}$ is the length of
the knot (four third power law)\cite{Cantarella,Buck}. For fluctuating, nonideal
polymers one can consider the average of $N_c$ over
all configurations. It is expected that quite generally,
for a compact SAP, $\langle N_c \rangle \sim N^{4/3}$~\cite{Grassberger}.
This result should indeed hold even for open walks
and be independend of topology for knotted SAP.
The fact that we find $\langle l \rangle \sim N $
for a fluctuating, knotted compact SAP suggests
that for real knotted polymer rings a
statistical generalization of the four third power law
of ideal knots should hold:
$\langle N_c \rangle \sim \langle l \rangle^{4/3}$.
In the case of ideal knots the growth of $l_{id}$
corresponds to an increase of the topological
complexity of the knot considered~\cite{Cantarella,Buck,Stasiak96}. 
For real knotted
rings the topology remains fixed with increasing $N$, while
fluctuations are able to produce the same type of  growth of
$\langle l \rangle$ versus $\langle N_c \rangle$.

In conclusion, we gave a consistent and robust statistical description
of the localization properties of prime knots in polymers. Weak 
localization and delocalization hold in the
good solvent and collapsed regimes, respectively.

This work was supported by  MIUR-COFIN03, FIRB01 and INFM-PAIS02.

%%%%%%%% FIG  5  %%%%%%%%%%%%%%%%%%%%%%%%%%%%%%%%%%%%%%%%%%%%%%%%%%%%%%%%%
\begin{figure}[!tbp]
\includegraphics[angle=0,width=\WIDTHA]{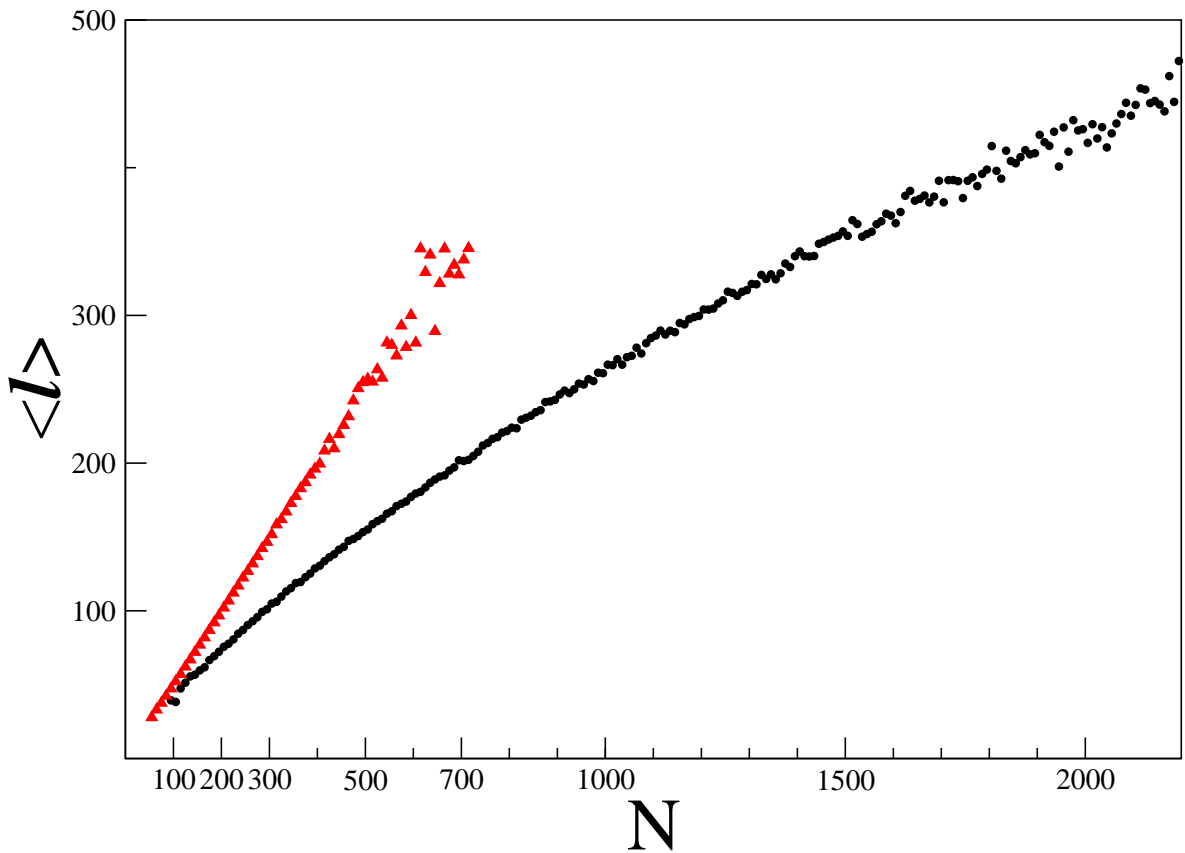}
\caption{ \label{fig5}
Plots of $\langle l \rangle $  of the shortest 
loop as a function of $N$. Both
curves correspond to the presence of one $3_1$ in each loop.
The bottom curve refers to   $T>>T_{\Theta}$ 
whereas the top to $T<T_{\Theta}$. 
}
\end{figure}


\begin{thebibliography}{35}
\bibitem{note0}
In fact the Alexander polynomial is not a perfect invariant, 
and is unable to distinguish every knot-type.
 For instance the prime knot $8_{11}$ has the
same Alexander polynomial as the composite knot $3_1\#6_1$, and $8_{15}$
has the same Alexander polynomial as $3_1\#7_2$.
To  distinguish pairs of knots we have also computed $\Delta (-2)$.
\bibitem{note1}
Similar procedures were used in Ref. [14,15] without addressing
asymptotic properties  like Eq. (1). 
For $N=500$ our algorithm misidentifies $\sim 5$ out of $1000$
unknotted SAP's, a much better rate than that reported in Ref. [14]
for phantom rings.
\end{thebibliography}
\end{document}